\documentclass[preprint,superscriptaddress,preprintnumbers,amsmath,amssymb,pra]{revtex4}
\usepackage{graphicx}
\usepackage{amsmath}

\begin{document}

\thispagestyle{empty}

\title{Reducing detrimental electrostatic effects in Casimir-force measurements
and Casimir-force-based microdevices }

\author{Jun~Xu}
\affiliation{Department of Physics and Astronomy, University of California, Riverside, California 92521, USA}

\author{
G.~L.~Klimchitskaya}
\affiliation{Central Astronomical Observatory at Pulkovo of the
Russian Academy of Sciences, Saint Petersburg,
196140, Russia}
\affiliation{Institute of Physics, Nanotechnology and
Telecommunications, Peter the Great Saint Petersburg
Polytechnic University, Saint Petersburg, 195251, Russia}

\author{
V.~M.~Mostepanenko}
\affiliation{Central Astronomical Observatory at Pulkovo of the
Russian Academy of Sciences, Saint Petersburg,
196140, Russia}
\affiliation{Institute of Physics, Nanotechnology and
Telecommunications, Peter the Great Saint Petersburg
Polytechnic University, Saint Petersburg, 195251, Russia}
\affiliation{Kazan Federal University, Kazan, 420008, Russia}

\author{
 U.~Mohideen\footnote{Umar.Mohideen@ucr.edu}}
\affiliation{Department of Physics and Astronomy, University of California, Riverside, California 92521, USA}

\begin{abstract}
It is well known that residual electrostatic forces create significant
difficulties to precise measurements of the Casimir force and to wide
use of Casimir-operated microdevices. We experimentally demonstrate
that with the help of Ar-ion cleaning of the surfaces it is possible
to make electrostatic effects negligibly small as compared to the
Casimir interaction. Our experimental setup consists of the dynamic
atomic force microscope supplemented with an Ar-ion gun and argon
reservoir. The residual potential difference between the Au-coated
surfaces of a sphere and a plate was measured both before and after
the in situ Ar-ion cleaning. It is shown that this cleaning decreases
the magnitude of the residual potential by up to an order of magnitude
and makes it almost independent on separation. The gradient of the
Casimir force was measured using ordinary samples subjected to the
Ar-ion cleaning. The obtained results are shown to be in good agreement
with both previous precision measurements using the specially selected
samples and with theoretical predictions of the Lifshitz theory. The
conclusion is made that the suggested method of in situ Ar-ion cleaning
is effective in reducing the electrostatic effects and therefore is
a great resource for experiments on measuring the Casimir interaction
and for Casimir-operated microdevices.
\end{abstract}

\maketitle

\section{Introduction}

During the last few years a great progress has been made in precision
measurements of the Casimir interaction caused by the zero-point and
thermal fluctuations of the electromagnetic field (see the most precision
experiments with conductor test bodies
\cite{23,24,25,25a,25b,26,27,28,29,30} and reviews \cite{36,37}).
This has raised a question on the possibility of using the Casimir
force in various micro- and nanoelectromechanical devices which find
expanding applications in fundamental science, engineering and
industry. As was understood decades ago \cite{1,2}, with shrinking
dimensions of microdevices to hundreds and tens nanometers the
fluctuation-induced van der Waals \cite{3} and Casimir \cite{4}
forces should make a pronounced effect on their functionality.
However, fifteen years have passed after publication of
Refs.~\cite{1,2} before  the first microdevices
driven by the Casimir force were demonstrated \cite{5,6}. This
gave impetus to research devoted to the role of stiction \cite{7,8},
electrostatic effects \cite{9}, surface roughness \cite{10,11,12},
geometry and dielectric properties of materials \cite{13,14}, and
phase transformations \cite{15} in nanotechnological systems
exploiting the Casimir force for their functionality. Various
possibilities to create the Casimir switch have also been
discussed \cite{16,17}. Finally, Casimir forces acting on a
micromechanical chip and between silicon nanostructures were
experimentally demonstrated \cite{18,19}. Hence, the engineering
and industrial applications of the Casimir microdevices are
within sight.

There is, however, a fundamental problem that plagued both the
investigation of the Casimir forces and their application in
nanodevices. It is connected with the existence of the residual
electrostatic forces acting between two grounded metallic bodies
in  vacuum. They originate from the surface patches due to  the
polycrystal structure of a metal, impurities, dust, surface
contaminants and resulting work-function inhomogeneities, and
may significantly distort the magnitude of the Casimir force.
In fact the electric forces caused by the distribution of
patch potentials depend critically on the surface preparation
and may at times be negligibly small and at other times relatively
large, as compared to the measured force (see Refs.~\cite{20,21}
and literature therein). They should be taken into account in
any measurement of small forces between the closely spaced
metallic surfaces, e.g., the gravitational force \cite{22}.

In measurements of the Casimir interaction the electrostatic
force due to surface patches manifests itself as a dependence
of the residual potential difference on the separation between
the surfaces. In the most precise experiments on measuring the
gradient of the Casimir force between a sphere and a plate (or
the effective Casimir pressure between two parallel plates) made
of Au \cite{23,24,25,26} or Au and Ni \cite{27,28,29} the residual
potential difference was demonstrated to be independent of
separation. This was achieved through the rejection of all
samples possessing the separation-dependent
residual potentials. Note
that in the difference measurement of the Casimir force \cite{30}
(experiments of this kind have been proposed in Refs.
\cite{31,32,33}) the role of surface patches was largely
eliminated. In fact a rejection of many of the prepared samples
is a plausible strategy when speaking about some unique fundamental
experiment. Such an approach is, however, not acceptable in the
day-to-day fabrication and operation of microdevices driven by the
Casimir force. Thus, the problem of surface patches in Casimir
physics calls for further investigation.

In this paper, we demonstrate the method allowing nearly full
elimination of the electrostatic effects in the Casimir-operated
microdevices. We use the experimental setup on measuring the
gradient of the Casimir force between two Au-coated surfaces of
a sphere and a plate by means of dynamic atomic force microscope
described in Refs. \cite{26,27,28,29}. The main modification we
made in this setup is that the vacuum chamber was equipped for
doing the cleaning of sphere and plate surfaces by means of argon
ion beam. It has been known that an interaction with a controlled
ionized Ar beam can be used for an improvement of the surface
quality including the removal of impurities and contaminants
\cite{34}. The ion beam cleaning is commonly used in the
manufactoring of electronic devices.
 According to our results, the multistep Ar-ion
cleaning of the ordinary (contaminated) plate and sphere surfaces
leads to up to an order of magnitude decrease of the residual
potential and significantly weakens its dependence on separation.
As a consequence, the electrostatic force becomes negligibly small
comparing to the Casimir force and not detrimental for the
functionality of a microdevice. We also perform measurements of
the gradient of the Casimir force and the effective Casimir
pressure between two parallel plates using the cleaned samples
and demonstrate good agreement with both experimental and
theoretical results of Refs.~\cite{23,24,25,26,27,28,29} obtained
with specially selected samples. Thus, the suggested method opens
up novel avenues in measuring the Casimir interaction and
for application of the Casimir force in
nanotechnology.

The paper is organized as follows. In Sec.~II we present the
brief description of the experimental setup. Section III
contains the experimental results and their comparison with
theory. In Sec.~IV the reader will find our conclusions and
discussion.

\section{The experimental setup}

The general scheme of the experimental setup is shown in Fig. 1.
It includes (1) the Ar-ion gun (RBD Model 04-165), (2) the
reservoir of Ar gas, and (3) the cantilever of an atomic
force microscope with attached Au-coated sphere of $R=60.8\mu$m
radius. The chip 4 holding the cantilever is mounted on two
piezoelectric actuators 5 and 6 using a clip. In the dynamic
regime employed in this experiment they are used to drive
oscillations of the cantilever at its resonant frequency with
a constant amplitude. The Au-coated plate is mounted on top of
a tube piezoelectric actuator 7 capable of travelling a distance
of $2.3\,\mu$m. The first fiber optic interferometer 8 monitores
the displacements of the Au plate mounted on the piezoelectric
actuator 7. The second fiber interferometer 9 records the
cantilever oscillations. The experiments were done in a
high-vacuum chamber with pressure down to $10^{-9}\,$Torr. For the
frequency demodulation the phase locked loop has been used.
All details of the experimental setup and its calibration are
described in Ref.~\cite{26}. Because of this, below we
concentrate only on the novel elements represented by the Ar-ion
cleaning system.

As mentioned in Sec.~I, the Ar-ion cleaning has long been used
for improvement of the quality of surfaces particularly
in surface science. It was applied by
many authors for different purposes as a method which does
not change an alignment and does not modify the structures in
the experiments \cite{34,38,39,40,41,42,43,44,45}. In our setup
(see Fig.~1) the Ar ion gun is mounted horizontally to the left
of the sphere and plate with a 2.75 inch CF flange. The distance
of the sphere to the output of the Ar-ion gun nose cone is
about 15\,cm. There is a clear line of sight from the Ar-ion gun
to both a sphere and a plate. The horizontal mounting of the
Ar-ion gun leads to the grazing incidence of Ar ions on the
plate surface. This reduces the sputtering of Au \cite{40}.
To reduce its outgassing rate, the ion gun was wrapped with a
heat tape and baked to over 100$\,{}^{\circ}$C for 24 hours. The source of
the Ar ions is from an Ar gas reservoir filled to 2\,atm of Ar
which is mounted below the gun using a 1.3 inch CF flange.

To operate the Ar-ion gun, the main vacuum chamber was
backfilled with Ar gas to a static pressure of
$2.5 \times 10^{-5}\,$Torr using the Ar gas reservoir.
The Ar atoms were ionized by
electron impact within the ion source's dual filament ionization
chamber. In the Ar-ion gun assembly, the dual tungsten filaments
were arranged off-axis to minimize the impurity content of the
Ar-ion beam. The ionized Ar ions were then extracted from the
ionization chamber, accelerated through the focus lens, and
directed to the sphere and the plate. To reduce the negative
impact of Ar ions, both the kinetic energy, as well as the ion
flux, were controlled. High ion kinetic energies lead to Au
atoms sputtering off the sphere and plate surfaces as well as
to burial of Ar ions due to their deep penetration \cite{38}.
The burial of Ar ions in the Au can be monitored as a change of
the residual potential when the ion gun is turned off. Various
ion kinetic energies from 2000 to 500\,eV were tried. The higher
ion kinetic energy leads to rapidly time varying residual
potential immediately after the Ar-ion cleaning due to the
aforementioned buried Ar ions in the Au.
As a result,
the Ar-ion kinetic energy of 500\,eV and
flux of $10\,\mu$A have been used. The different Ar-ion
cleaning times have been probed under these conditions.

The extent of contamination of Au surfaces as a function of the
Ar-ion cleaning time was monitored through the change in the
residual potential difference. This is essentially equivalent to
the commonly used surface science probes such as the Auger
electron spectroscopy and X-ray photoelectron spectroscopy.

\section{The experimental results and comparison with theory}

The first measurements have been made with an ordinary (uncleaned)
sphere and plate exposed during some time to an ambient
environment. These data were taken using the procedure described
in detail in Ref.~\cite{26}. Specifically, the absolute separations
were determined as $a=a_0+z_{\rm piezo}$, where $z_{\rm piezo}$
is the distance traveled by the plate due to the voltage applied
to the piezoelectric actuator situated under the plate,
and $a_0$ is the closest separation between the Au sphere and
Au plate. Then 11 different voltages $V_i$ have been applied
to the plate and the corresponding frequency shift
$\Delta\omega_i$ have been measured. Any drift of the frequency
shift signal due to the mechanical drift of the piezo actuator
was subtracted using the procedure described in Ref.~\cite{26}.

The obtained frequency shift $\Delta\omega$ has a parabolic
dependence on the applied voltage \cite{26}
\begin{equation}
\Delta\omega=-\beta(V-V_0)^2-C\frac{\partial F_C}{\partial a},
\label{eq1}
\end{equation}
\noindent
where $F_C$ is the Casimir force, $C=\omega_0/(2k)$ is expressed via
the natural frequency of the oscillator $\omega_0$ and the spring
constant of the cantilever $k$, and the coefficient $\beta$ is given by
\begin{eqnarray}
&&
\beta=2\pi\epsilon_0C\sqrt{\frac{1}{a(2R+a)}}\sum_{n=1}^{\infty}
{\rm csch}(n\alpha)
\label{eq2} \\
&&~~~~~~~\times
\left\{
n\coth(n\alpha)[n\coth(n\alpha)-\coth\alpha]\right.
\nonumber \\
&&~~~~~~~~~~~~~~~~~~~\left.
-{\rm csch}^2\alpha+n^2{\rm csch}^2(n\alpha)\right\}.
\nonumber
\end{eqnarray}
\noindent
Here, $\epsilon_0$ is the permittivity of a free space,
$\cosh\alpha\equiv 1+a/R$, and $R$ is the sphere radius.
Equation (\ref{eq2}) is obtained by differentiation with respect
to separation of the exact expression for the electrostatic force
in the sphere-plate geometry \cite{26,4,38a}.

The residual potential difference $V_0$, corresponding to the
position of the parabola maximum, was found from Eq.~(\ref{eq1})
using $\chi^2$ fitting procedure over the entire measurement range
from $a_0$ to 700\,nm.
In a similar way the data for the curvature of the parabola $\beta$
were obtained.
The quantities $a_0$ and $C$ were found by fitting the data for
$\beta$ as function of the separation distance to the theoretical
expression (\ref{eq2}).

In Fig.~\ref{fg2} the resulting values of the residual potential
difference are shown as dots at each absolute separation from 235
to 700\,nm with a step of 1\,nm. As is seen in Fig.~\ref{fg2},
for an uncleaned sample $V_0$ takes rather high values exceeding
32\,mV. In Fig.~\ref{fg2} we also show the best fit of the obtained
values of $V_0$ to the straight line leaving its slope and an
initial point as free parametrs. This line is described by
the equation
\begin{equation}
V_0=(2.60\times 10^{-3}a+31.95)\,\mbox{mV},
\label{eq3}
\end{equation}
\noindent
where $a$ is measured in nm.
{}From Fig.~\ref{fg2} one can conclude that $V_0$ substantially depends
on separation.
The change of $V_0$ between final and initial points of the straight line
is equal to approximately 1.3\,mV.

Next we performed several measurements of $V_0$ at the closest separation
in order to demonstrate the effect of Ar-ion cleaning qualitatively.
To simplify the procedure, in these measurements the small corrections
due to mechanical drift of the piezo actuator have been disregarded.
The initial measurement performed before starting the cleaning
resulted in $V_0=31.8\,$mV. Then we did the Ar-ion cleaning of the
sphere and plate surfaces during 5\,min and again determined the
value of the residual potential difference at the closest separation
$V_0=20.5\,$mV. After that we have performed three more Ar-ion cleaning
procedures, with the durations 10, 20, and 63\,min,
and after each of them determined the
value of $V_0$. Thus, after the total pariod of 98\,min of
Ar-ion cleaning the value of $V_0=2.4\,$mV was obtained.
In Fig.~\ref{fg3} the residual potential diffrence is plotted
as a function of the cleaning time. As is seen in this figure, for
smaller $V_0$ its value decreases more slowly with the cleaning time.
After 98\,min of the Ar-ion cleaning
we observe a decrease of $V_0$ by the factor of 13.

For a full quantitative investigation of the effect of Ar-ion cleaning
on $V_0$ we have done additional Ar-ion cleaning of the surface for 100\,min.
Thereafter the accurate measurements of $V_0$ for cleaned surfaces
as a function of separation have been performed  with a step of 1\,nm
taking into account the drift of the piezo actuator. The obtained results
are shown as dots in Fig.~\ref{fg4}(a). As is seen in this figure,
an additional Ar-ion cleaning resulted in further decrease of $V_0$
(up to a factor of 30 at the shortest separation as compared with
the case of an uncleaned sample in Fig.~\ref{fg2}).
The best fit of the data for $V_0$ to the straight line, which is
also shown in Fig.~\ref{fg4}(a), results in
\begin{equation}
V_0=(1.07\times 10^{-3}a+0.928)\,\mbox{mV}.
\label{eq4}
\end{equation}
\noindent
{}From the comparison of this equation with Eq.~(\ref{eq3}) one finds
that the slope of the straight line decreased by a factor of 2.43.
The respective drop of $V_0$ between final and initial separation
distances is equal to 0.50\,mV, i.e., became smaller by a factor
of 2.60.

After finishing of this measurement, we have performed one more
Ar-ion cleaning of the surface with a duration of 60\,min.
Then the residual potential difference was measured again as
a function of separation. The measurement results
are shown as dots in Fig.~\ref{fg4}(b) as well as the
best fit  to the straight line
\begin{equation}
V_0=(0.917\times 10^{-3}a-5.80)\,\mbox{mV}.
\label{eq5}
\end{equation}
\noindent
As a result of one more Ar-ion cleaning, the slope of this
line has decreased further [by a factor of 2.84 in comparison
with an uncleaned sample in Eq.~(\ref{eq3}) and by a factor of
1.17 compared to  Eq.~(\ref{eq4})]. Note also that the
dispersion of dots in Fig.~\ref{fg4}(b) is much smaller than
in Fig.~\ref{fg4}(a).

Finally, we have determined the experimental values of the gradient
of the Casimir force for the measurement set of Fig.~\ref{fg4}(b)
obtained after the longest Ar-ion cleaning. This was done using
Eq.~(\ref{eq1}) where the applied voltage $V$ was put
equal to the mean residual voltage
$\langle{V}_0\rangle=-5.37\,$mV as is
in Eq.~(\ref{eq5}) and Fig.~\ref{fg4}(b).
At each separation the voltage $V=\langle{V}_0\rangle$ was applied
11 times and the mean value of the gradient of the Casimir force
was obtained. Then it was recalculated into the effective
Casimir pressure between two parallel Au plates using the proximity
force approximation \cite{4}
\begin{equation}
P_C(a)=-\frac{1}{2\pi R}\,\frac{\partial F_C(a)}{\partial a},
\label{eq6}
\end{equation}
\noindent
which is sufficiently precise for our purposes.
As was shown recently \cite{38b,38c,38d}, the relative correction
to the proximity force approximation is approximately equal to
$0.5a/R$, i.e., varies from 0.19\% to 0.33\% when the separation
 increases from 235 to 400\,nm.
In Fig.~\ref{fg5} the obtained values of the Casimir pressure
are indicated as crosses over the separation regions (a) from
235 to 350\,nm with a step of 1\,nm and (b) from 350 to 700\,nm
with a step of 3\,nm for better visualization. The arms of the
crosses show the total experimental errors determined at a 67\%
confidence level. The absolute error in the measurements of
separation is equal to 0.5\,nm. The relative error in the
Casimir pressures at $a = 235\,$nm is equal to approximately 2\%
and increases with increasing separation.
Thus, it is  an order of magnitude larger than that  introduced
from using the proximity force approximation.
Note that in
Ref.~\cite{26}, where all measurements have been performed using
selected samples which have $V_0$ independent of
separation, the error in the Casimir pressure at the
shortest separation was equal to 0.75\%.

We have also compared the measured values of the effective
Casimir pressures with theoretical predictions of the Lifshitz
theory computed using the tabulated optical data for the
complex index of refraction of Au \cite{35}. It has long been
known that there is the outstanding problem in the comparison
of the Lifshitz theory with experiment connected with an
extrapolation of the available optical data to zero frequency
by either the Drude or the plasma model
(see, e.g., Refs.~\cite{4,36,37}).
The point is that all precise experiments on
measuring the Casimir interaction between metallic test bodies
exclude the predictions of the Lifshitz theory if this
extrapolation is made by the Drude model taking into account
the relaxation properties of conduction electrons
\cite{23,24,25,26,27,28,29,30}.
The same experiments are in a very good
agreement with the theoretical predictions when the extrapolation
is made by the lossless plasma model. This result has been
conclusively confirmed by the differential force measurement of
Ref.~\cite{30}, where the difference in theoretical predictions
using two different extrapolations was up to a factor of $10^{3}$.

We have computed the Casimir pressure between two parallel plates
made of Au at the laboratory temperature $T = 300\,$K using the
Drude and plasma extrapolations of the optical data to zero
frequency in the same way as in Ref.~\cite{26}. The obtained
results are shown in Fig.~\ref{fg5} by the upper and lower solid lines,
respectively. As is seen in this figure, the predictions of the
Lifshitz theory using the plasma model for an extrapolation
of the optical data (the lower line) are in a very good agreement
with the measurement data over the entire measurement range. The
theoretical predictions using an extrapolation by means of the
Drude model (the upper line) are excluded by the data over the
separation region from 235 to 400\,nm. Thus, the use of ordinary
(not especially selected) a sphere and a plate, which were
subjected to the Ar-ion cleaning, leads to qualitatively the
same conclusions concerning a comparison between experiment and
theory as the most precise experiments performed previously.
The drawback of using the Ar-ion cleaned ordinary samples lies
in the somewhat larger measurement errors, but this point is of
more importance for fundamental research rather than for the
manufacture and functionality of microdevices.

Taking into account that in Fig. 5(b) some dependence of $V_0$
on separation still persists, it is interesting to compare the
remaining noncompensated electric force with the Casimir force.
Calculations using Eq. (4) show that at
separations $a = 235$, 300, 400, and
700\,nm the electric pressure constitutes  only 0.9, 1.3, 2.1,
and 5.2\% of the Casimir pressure, respectively. Thus, one
can neglect the role of residual electric forces when
considering the functionality of microdevices using the Ar-ion
cleaned samples. Quite to the contrary, before the Ar-ion cleaning
it was
$\langle{V}_0\rangle = 33.16\,$mV (see Fig. 2) and the noncompensated
electric force was equal to 30, 46, 77, and 222\% of the Casimir
pressure at the same respective separations. This means that the
use of regular samples with no cleaning for manufacturing the
microdevices driven by the Casimir force
and for measuring the Casimir interaction
is seriously hindered
by electric forces due to the surface patches, contaminants etc.

\section{Conclusions and discussion}

In the foregoing, we have investigated the impact of Ar-ion
cleaning on the electrostatic effects arising between two closely
spaced metallic surfaces in measurements of the Casimir
interaction. For this purpose, the previously created experimental
setup using the dynamic atomic force microscope was supplemented
with an Ar-ion gun and the Ar reservoir. Our experimental results
demonstrate that when using regular uncleaned surfaces of a
sphere and a plate the residual potential difference between these
surfaces is rather large in magnitude and depends considerably
on the separation distance, as it was repeatedly reported in
previous literature (see Ref.~\cite{21} and references therein).
We have also shown that sufficiently long Ar-ion cleaning diminishes
the residual potential difference by up to an order of magnitude and
makes it less dependent on separation. Thus, for uncleaned surfaces
the residual electric pressures were determined to vary from 30\%
to 222\% of the Casimir pressure when separation varies from 235
to 700\,nm. This is a serious obstacle for a regular application
of microdevices driven by the Casimir force
and for measuring the Casimir interaction. By contrast, after
the Ar-ion cleaning the residual electric pressures vary from
only 0.9\% to 5.2\% of the Casimir pressure in the same range
of separations which leaves the Casimir pressure  as the
dominant driver.

We have also measured the gradients of the Casimir force between
Au surfaces of a sphere and a plate and determined the respective
Casimir pressures between two parallel plates for the Ar-ion
cleaned samples. The obtained results have been compared with
the previously performed most precise measurements using
specially selected clean samples and with theoretical
predictions of the Lifshitz theory. It was found that, although
the measurement data from the Ar-ion cleaned samples are
characterized by somewhat larger experimental errors than that of the most
precise experiments, it is in agreement with the latter and with
the Lifshitz theory using the plasma model for extrapolation of
the optical data to zero frequency. The theoretical predictions
of the Lifshitz theory using the Drude model for the extrapolation
of the optical data are excluded by our measurements with the
Ar-ion cleaned samples over a wide separation region from 235
to 400 nm.

One can conclude that the method of in situ Ar-ion cleaning of
the interacting surfaces
is effective in reducing the residual potential difference and
related electrostatic effects and therefore is a great resource
 for measuring the Casimir interaction and the wide
use of microdevices operated by the Casimir force.

\section*{Acknowledgments}

The work of J.X.~and U.M.~was partially supported by the NSF
grant PHY-1607749.
The work of V.M.M.{\ }was partially supported by the Russian
Government
Program of Competitive Growth of Kazan Federal University.


\newpage
\begin{widetext}
\begin{figure}[b]
\vspace*{-4cm}
\centerline{\hspace*{2.cm}
\includegraphics{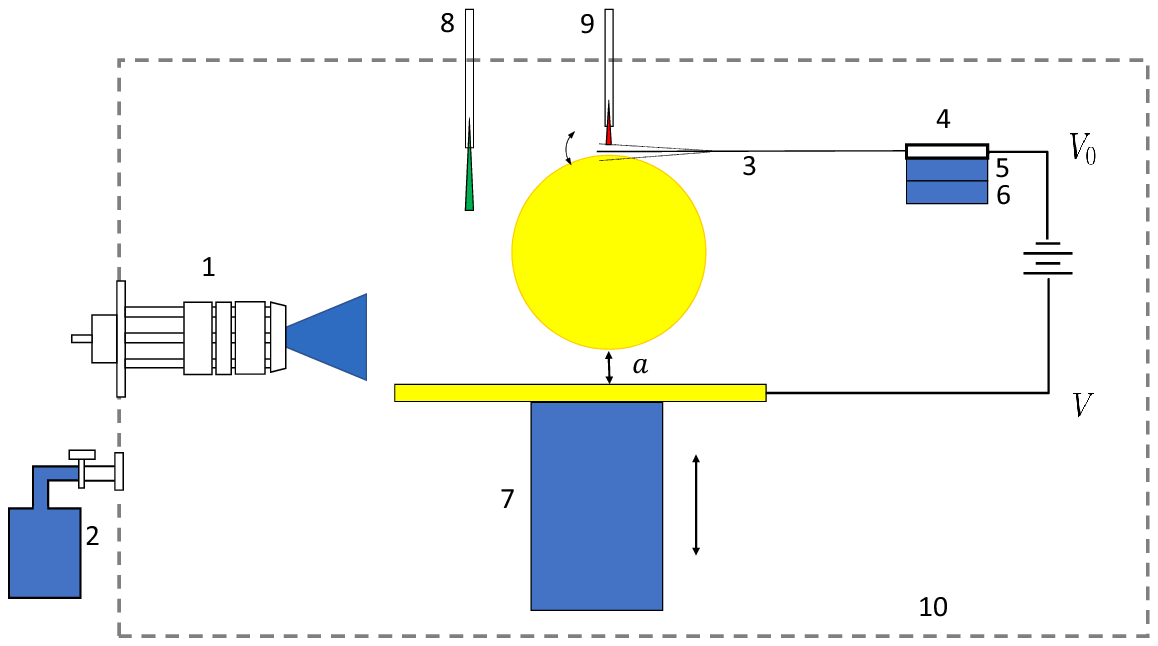}
}
\vspace*{-10cm}
\caption{\label{fg1}
Schematic diagram of the experimental setup: 1 is the
Ar-ion gun, 2 is the reservoir of Ar gas, 3 is the cantilever
of an atomic force microscope with an attached sphere, 4 is the
cantilever holder, 5, 6, and 7 are the piezoelectric actuators
of this holder and of the plate, 8 and 9 are the fiber optic
interferometers, and 10 is the vacuum chamber.
}
\end{figure}
\begin{figure}[b]
\vspace*{-4cm}
\centerline{\hspace*{2.5cm}
\includegraphics{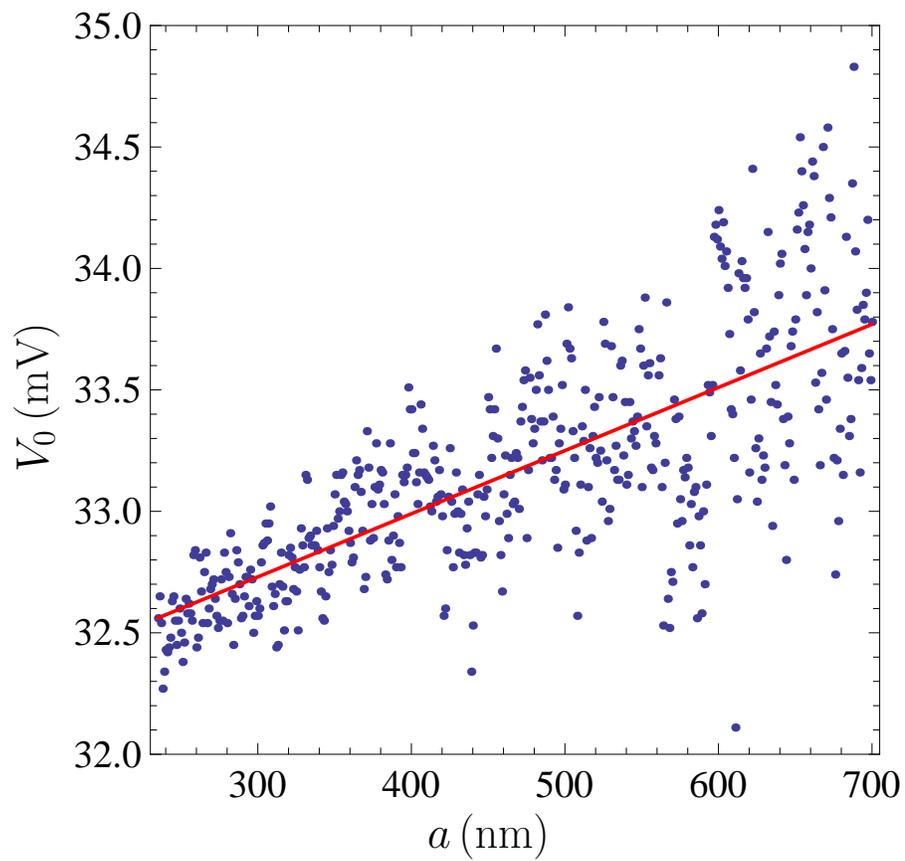}
}
\vspace*{-10cm}
\caption{\label{fg2}
The residual potential difference between the uncleaned
a sphere and plate is shown by dots as a function of
separation. The best fit of $V_0$ to a straight line is
also shown.}
\end{figure}
\begin{figure}[b]
\vspace*{-4cm}
\centerline{\hspace*{2.5cm}
\includegraphics{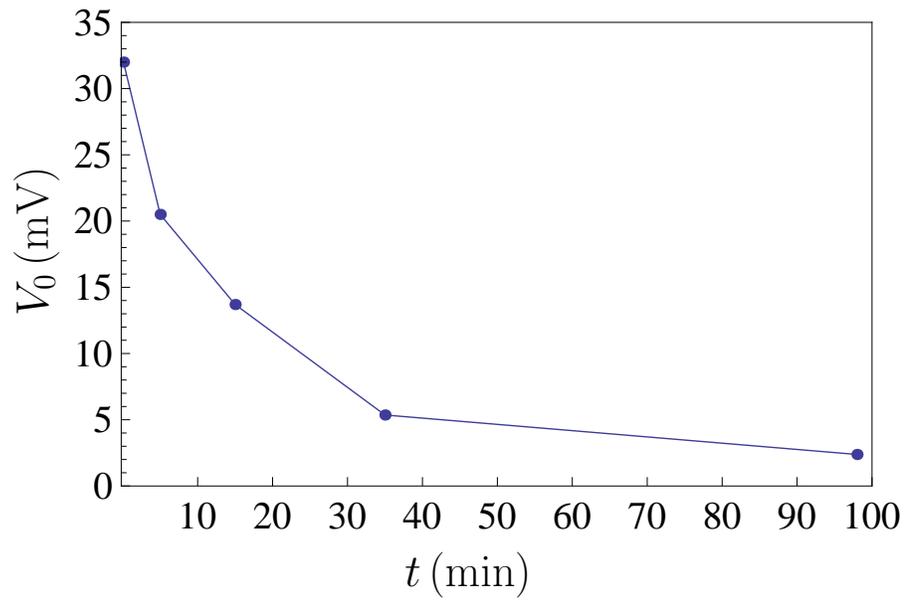}
}
\vspace*{-10cm}
\caption{\label{fg3}
The residual potential difference between a sphere
and a plate is shown by dots as a function of duration of
the Ar-ion cleaning.
}
\end{figure}
\begin{figure}[b]
\vspace*{1cm}
\centerline{\hspace*{2.5cm}
\includegraphics{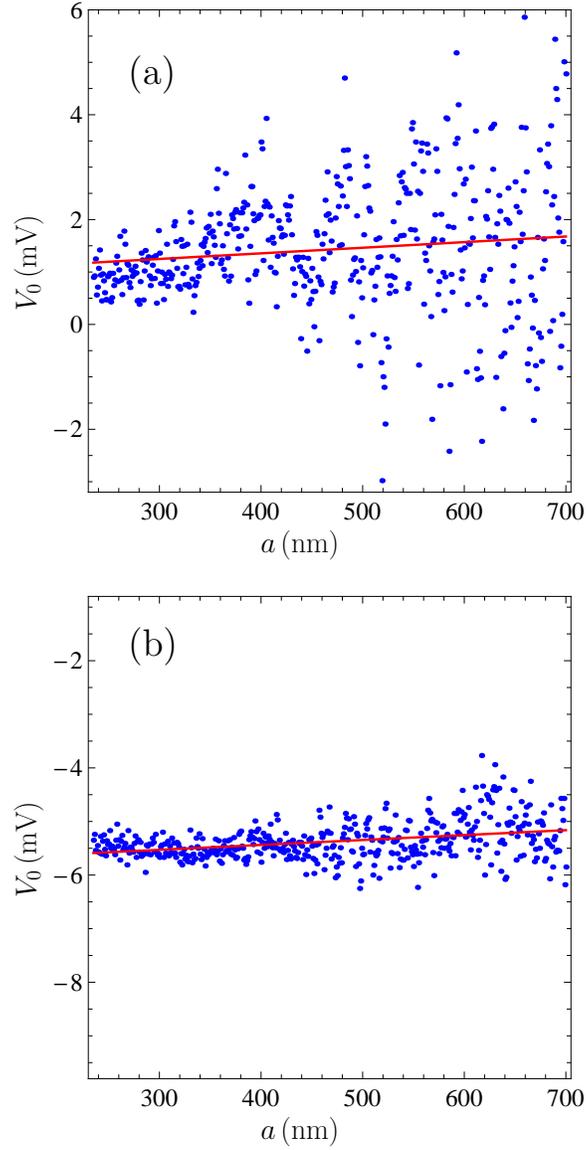}
}
\vspace*{-15cm}
\caption{\label{fg4}
The residual potential difference between a sphere
and a plate is shown by dots as a function of separation
after (a) additional Ar-ion cleaning of 100 min duration and
(b) additional Ar-ion cleaning of 160 min duration. In both
cases the best fit of $V_0$ to a straight line is
also shown.}
\end{figure}
\begin{figure}[b]
\vspace*{1cm}
\centerline{\hspace*{2.5cm}
\includegraphics{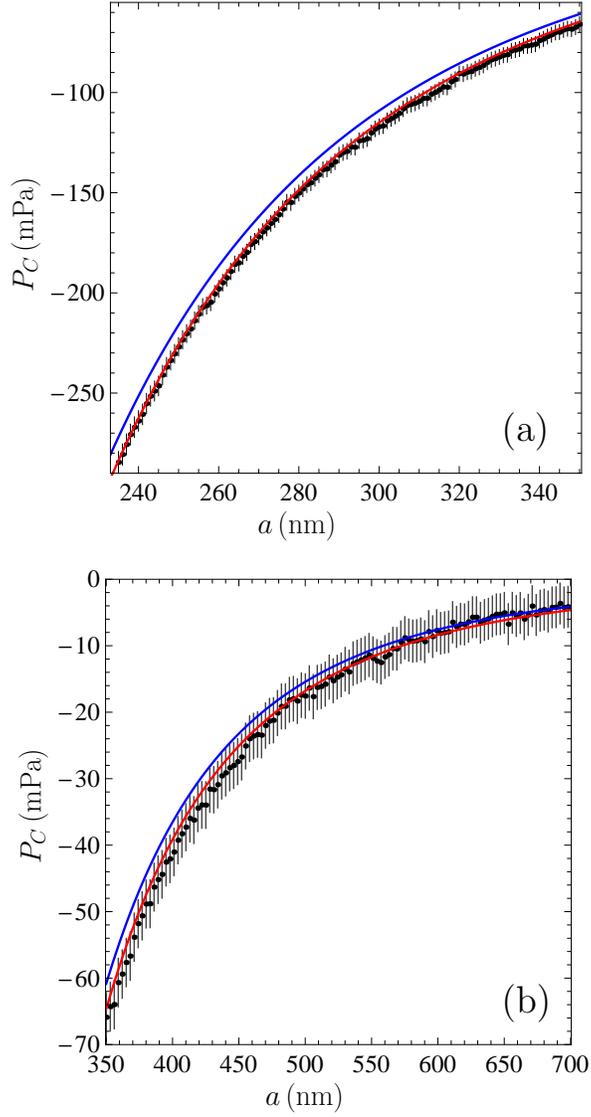}
}
\vspace*{-15cm}
\caption{\label{fg5}
Comparison between the mean measured Casimir pressures
(crosses plotted at a 67\% confidence level) and theory (upper
and lower lines computed using the Drude and plasma model
extrapolations of the optical data, respectively) within the
separation regions (a) from 235 to 350\,nm and (b) from 350 to
700\,nm.}
\end{figure}
\end{widetext}
\end{document}